\documentclass[twocolumn,english]{article}

\usepackage[T1]{fontenc}
\usepackage[latin9]{inputenc}
\usepackage{verbatim}
\usepackage{float}
\usepackage{amssymb}
\usepackage{graphicx}
\usepackage{amsmath}

\usepackage{fancyhdr}

\usepackage{hyperref}

\makeatletter

\floatstyle{ruled}
\newfloat{algorithm}{tbp}{loa}
\providecommand{\algorithmname}{Algorithm}
\floatname{algorithm}{\protect\algorithmname}

\newcommand{\lyxaddress}[1]{
\par {\raggedright #1
\vspace{1.4em}
\noindent\par}
}

\@ifundefined{definecolor}
 {\usepackage{color}}{}
\@ifundefined{definecolor}
 {\@ifundefined{definecolor}
 {\usepackage{color}}{}
}{}
\usepackage{babel}

\usepackage{babel}\usepackage{amsfonts}\@ifundefined{definecolor}
 {\@ifundefined{definecolor}
 {\usepackage{color}}{}
}{}
\usepackage{array}\usepackage{supertabular}\usepackage{hhline}

\newcommand{\arraybslash}{\let\\\@arraycr}

\newcommand{\ps@Standard}{
  \renewcommand\@oddhead{}
  \renewcommand\@evenhead{}
  \renewcommand\@oddfoot{}
  \renewcommand\@evenfoot{}
  \renewcommand\thepage{\arabic{page}}
}

\usepackage{babel}

\newcommand{\sandwich}[3]{\left \langle #1 \mid #2 \mid #3 \right\rangle}

\@ifundefined{showcaptionsetup}{}{%
 \PassOptionsToPackage{caption=false}{subfig}}


\usepackage{babel}
\usepackage[normalem]{ulem}

\usepackage{algorithm,algpseudocode}

\makeatother

\usepackage{babel}

\begin{document}

\title{Modernizing Quantum Annealing II: Genetic algorithms with the Inference Primitive Formalism}

\author{Nicholas Chancellor\thanks{email: nicholas.chancellor@durham.ac.uk}}

\maketitle

\lyxaddress{Department of Physics, Durham University, South Road, Durham, UK}
\begin{abstract}

Quantum annealing allows for quantum fluctuations to be used used to assist in finding the solution to some of the worlds most challenging computational problems. Recently, this field has attracted much interest because of the construction of large-scale flux-qubit based quantum annealing devices. There has been recent work on [Chancellor NJP 19(2):023024, 2017] how the control protocols of these devices can be modified so that individual annealer calls on real devices can take initial conditions. Development is being undertaken to implement such protocols in the quantum annealing devices designed by D-Wave Systems Inc. and these features will be available to customers soon. In this paper, I develop a formalism for algorithmic design in quantum annealers, which I call the `inference primitive' formalism. This formalism allows for a natural description of calls to quantum annealers with a general control structure. This more generalized control structure includes not only the ability to include initial conditions in an annealer run, but also to control the annealing schedules of qubits or clusters of qubits independently, thereby representing relative certainty values of different parts of a candidate solution. I discuss the compatibility of such controls with a wide variety of other current efforts to improve the performance of annealers, such as non-stoquastic drivers, synchronizing freeze times for the qubits, and belief propagation techniques. To demonstrate the power of the formalism I present here, I discuss how this new formalism can be used to represent annealer implementations of genetic algorithms, and can represent the addition of genetic components to currently used algorithms. The new tools I develop will allow a more complete understanding of the algorithmic space available to quantum annealers, and thereby make the field more competitive.

\end{abstract}

\thispagestyle{empty}

\section{Introduction}

The quantum annealing algorithm (QAA) \cite{Finella1994,kadowaki1998,Kaminsky2002, Kaminsky2004, Kaminsky2004a} has been demonstrated to be a promising candidate for a vast number of real-world problems. The potential applications are too numerous to list here, but include fields as diverse as aerospace \cite{Coxson2014}, computational biology \cite{perdomo-ortiz2012}, neural networks \cite{Amin2016,Benedetti2016,Benedetti2016a,Adachi2015}, pure computer science \cite{Chancellor/Zohren2016}, and economics \cite{Marzek}. In this manuscript, I discuss a formalism which can represent general control of quantum annealers. I demonstrate how this formalism can be used to design new algorithms based on multiple calls to a quantum annealer. More generally, this formalism represents hybrid analog-digital computation, but I restrict the discussion in this paper to quantum annealing applications. 

The QAA as it is usually structured starts from a superposition state representing all possible solutions. The system is then annealed and quantum fluctuations are introduced through competition between a problem Hamiltonian and a `driver' Hamiltonian which does not commute with the problem Hamiltonian
\begin{equation}
H(s)=A(s(t))\,H_{\text{driver}}+B(s(t))H_{\text{problem}}, \label{eq:H_Anneal}
\end{equation}
where $0\le s \le 1$ is the annealing parameter which controls the annealing schedule, $A(s(t))$, $B(s(t))$, which are chosen such that $\frac{A(0)}{B(0)}\gg1$ and $\frac{B(1)}{A(1)}\gg1$, and both behave monotonically with $s$. In traditionally formulated quantum annealing, $s$ is also a monotonic function of $t$, but to construct the protocols here, I will consider cases where $s$ is a non-monotonic function of $t$, as was done in \cite{Chancellor2016}. The problem Hamiltonian is usually chosen to be an Ising model,
\begin{equation}
H_{Problem}=-\sum_{i}h_{i}\sigma_{i}^{z}-\sum_{i,j\in\chi}J_{ij}\sigma_{i}^{z}\sigma_{j}^{z},\label{eq:ISGham}
\end{equation}
 with field variables $h_i$ and coupler variables $J_{ij}$. Ising model-based annealing architectures were first proposed in the context of closed quantum systems by Kadowaki and Nishimori \cite{kadowaki1998} and later generalized to open quantum systems by Kaminski, Lloyd and Orlando \cite{Kaminsky2002, Kaminsky2004, Kaminsky2004a}. In this paper I consider open system quantum annealing, where tunneling mediated by these fluctuations is driven by a low temperature thermal bath. One example of a driver Hamiltonian is the transverse field driver which is currently implemented on the annealers produced by D-Wave Systems Inc.~\cite{D-Wave}. 
\begin{equation}
H_{driver}=-\sum_{i}\sigma_{i}^{x}\label{eq:transverseIsing}
\end{equation}
 I also consider more general multi-body driver Hamiltonians of the form 
\begin{equation}
H_{driver}=\sum_{i}c_i \prod_{j \in R_i}\sigma_{j}^{(\phi_i)}\label{eq:multi-bodyDrive}
\end{equation}
where, $c_i$ is a positive real number which determines the strength of the coupling, $R_i$  is a set of qubits, and
\[
\sigma_{j}^{(\phi)}=(\exp(i \, \phi)\,a_{j}+\exp(-i \, \phi)\,a_{j}^{\dagger}),
\]
where $a=\left(\begin{array}{cc} 0 & 1 \\ 0 & 0 \end{array} \right)$ is a lowering operator operator such that $\sigma^x=a+a^\dagger$. The reason such drivers are of interest is that they are able to introduce a sign problem in quantum Monte Carlo simulations if no basis exists for which all off diagonal terms are negative \cite{Bravyi2008,Bravyi2009}. No other method is known for large scale low temperature simulations of these so-called non-stoquastic Hamiltonians \cite{Bravyi2014}. Because of this increased difficulty in simulation, it is widely suspected that quantum annealing with non-stoquastic drivers is more powerful than quantum annealing with stoquastic drivers.

Recall that the QAA as it is usually formulated starts from an equal superposition of all classical solutions, meaning that there is no way to incorporate existing knowledge about the solution, neither from previous annealing runs nor from different algorithms.  One way around this deficiency is to use algorithms based on local searches \cite{Chancellor2016,Amin_patent} around a candidate solution rather than global searches which start from a superposition of all classical solutions. In particular, \cite{Chancellor2016} includes proof-of-principle numerical experiments which demonstrated how such techniques may assist in a search. It has recently been announced that reverse annealing features capable of performing these protocols will be added to D-Wave 2000Q devices \cite{rev_anneal_announce}.

There is also an alternate formulation which pre-dates the proposals in \cite{Chancellor2016,Amin_patent}  which allows an initial guess \cite{Perdomo-Ortiz2011} to be incorporated into a closed system adiabatic quantum protocol. While protocols based on these techniques can also be represented with the inference primitive formalism, for this paper I will restrict the discussion to the local search formulation in \cite{Chancellor2016}. It also may be fruitful to explore connections to recent work exploring the use of a reinforcement algorithm \cite{Ramezanpour2017} in quantum optimisation, although such a study is beyond the scope of this work.

In addition to representing the protocols in \cite{Chancellor2016}, I show that the formalism demonstrated here represents a more generalized control strategy which includes annealing the qubits independently. Such additional freedom allows for the annealer to accept individual uncertainty values for each bit, or cluster of bits in the case of multi-body drivers. 

This formalism can be used to demonstrate a new way in which a directed mutation engine for genetic algorithms \cite{macKaybookGA,Deng1999,Fogel1994} can be constructed using these individual uncertainty values.  The idea of using an annealer for genetic algorithms is not new: Coxson, Hill, and Russo \cite{Coxson2014} experimentally demonstrated that a D-Wave device can successfully aid these algorithms in finding optimal radar waveforms. The method I propose, however, is completely general, and only requires that an annealer be able to realize a problem Hamiltonian, rather than a potentially more complex directed mutation Hamiltonian.

The structure of this paper is as follows. In Sec.~\ref{sec:inf_prim} I discuss the inference primitive formalism, how it relates to quantum annealers, and demonstrate how previously known algorithms such as the traditional QAA and those proposed in \cite{Chancellor2016} may be represented using inference primitives. In Sec.~\ref{alg_design} I discuss how annealer based genetic algorithms may be represented in this formalism and how it may be used to add genetic components to the algorithms proposed in \cite{Chancellor2016}.  This is followed by a discussion in Sec.~\ref{method_comp}  about how the control represented in the inference primitive formalism is compatible many other recent advances in the field, including synchronization of freezing \ref{sub:protocol}, higher order drivers, including non-stoquastic drivers \ref{sub:high_order}, and belief propagation methods used to represent graphs larger than the hardware\ref{sub:belief_propagation}.  Finally in Sec. \ref{conclusions} I conclude with some overall discussion.

\section{Inference Primitive \label{sec:inf_prim}}

Consider a high level description of a subroutine $\Phi$ which performs a guided search of an energy landscape based on known information about likely solutions. I will call such a subroutine an inference primitive, as it will try to infer the correct solution based on input information. The inference primitive will be supplemented by information processing which determines the parameters to give each call to the primitive, I will call this the processing function $\mathcal{F}$. I will demonstrate later in this section that $\Phi$ can be a high level description of a call to a quantum annealer, with $\mathcal{F}$ representing classical information processing used within a hybrid algorithms. I will also formally define both $\Phi$ and $\mathcal{F}$.

\begin{figure}
\begin{centering}
\includegraphics[width=7cm]{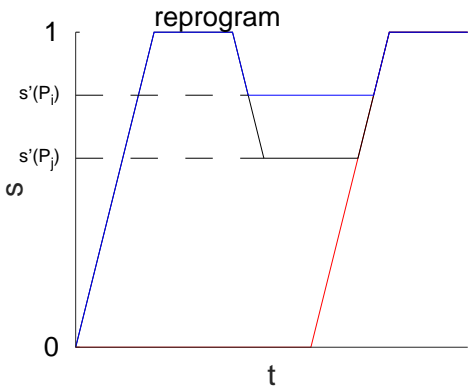}
\par\end{centering}
\caption{\label{fig:schedule} Annealing schedule for inference primitive protocol. This is the same as in \cite{Chancellor2016} except that  individual qubits are annealed back to different values of $s$. Qubits are annealed first with a simple Hamiltonian to program an initial state, then the Hamiltonian is reprogrammed to the problem Hamiltonian and each qubit (or multi-qubit driver) is annealed back to $s'(P_i)$, where $P_i$ is a measure of the uncertainty of a qubit value. The qubits are then annealed back toward $s=1$, each starting its anneal when the other bits reach the same value of $s$. For $s_i'=0$ (red), setting the initial value is unnecessary, as no information about the qubit value is known.}
\end{figure}

Before discussing the formalism further, I will motivate the use of this formalism to represent control of quantum annealers. It has recently been demonstrated in \cite{Chancellor2016}, that global transverse fields can be used to control the range of local search in solution space. Building on this idea, application of different transverse fields locally will cause an algorithm to search different ranges in different directions in solution space. In this way, the strength of local transverse fields can encode bitwise certainty of a solution. In fact, algorithms based on an extreme version of this have already been implemented  \cite{Karimi2017,Karimi2017a}, in which, based on previous solution statistics, qubits are either treated as taking fixed values (absolute certainty), or annealed using a traditional protocol (absolute uncertainty). To implement a protocol which incorporates local uncertainty, I generalize the methods given in \cite{Chancellor2016}, to allow different qubits to be annealed to different points $s_i'$, as depicted in Fig. \ref{fig:schedule}.

In this paper, I will not focus on how to construct heuristics which relate uncertainty to transverse field strengths, but rather examine how algorithms can be designed and represented, assuming a suitable heuristic has been developed. I provide an example of a very simple heuristic in appendix 1. This heuristic is only intended as an example of how these quantities can be related, and may be too simplified to perform well in the real world. Alternative heuristics could be based on experimental local temperature estimates  using the methods of \cite{Raymond2016}, or by adaptations of the methods to estimate a global effective temperature used in \cite{Benedetti2016}. For the remainder of this work, I will assume that a suitable heuristic, $s_i'(\{P\})$, where the set notation has been used to emphasize that in general this parameter may also depend on the uncertainty $P_i \in [0,0.5]$ of neighbouring qubits as well. 

\begin{figure}
\begin{centering}
\includegraphics[width=7cm]{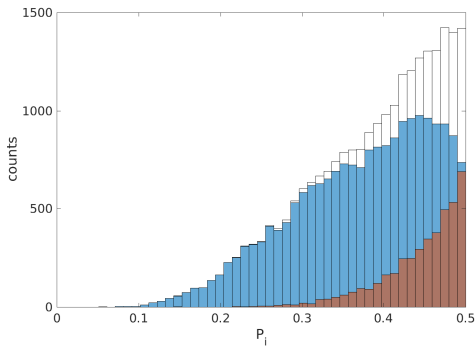}
\par\end{centering}
\caption{\label{fig:counts_P} Historgam of $P_i$ for spin values obtained by the `traditional' QAA  on $1500$ instances of spin glass problems described by Eq. (\ref{eq:SK_fix}) with $n=17$, ($1500\times (n-1)=24,000$ data points). Data are based on PIQA runs with $T=0.8246$ and $\tau=20$ using the same numerical methods as the proof of principle in \cite{Chancellor2016}. Blue bars are cases where $S_i$ found by Eq. (\ref{eq:S_raw}) agrees with the true ground state, red are cases where it does not, and unfilled bars are total counts.}
\end{figure}

I have motivated the high level description of a quantum annealer as an inference primitive $\Phi$, now I must further motivate that suitably chosen processing functions $\mathcal{F}$ will be able to appropriately extract uncertainty information from the output data of a quantum annealer. To do this, I consider the problem of finding the ground state of a Sherrington-Kirkpatrick like spin glass \cite{Sherrington1975}:
\begin{equation}
H_{SK}=-\sum_{i<j}^nJ_{ij}\sigma_i^z \sigma_j^z,
\end{equation}
where each $J_{ij}$ is selected uniformly randomly from the range $[-1,1]$. All energy eigenstates of such Hamiltonians will be at least two fold degenerate because of total spin inversion symmetry. To break this symmetry I fix the last spin to be in the down orientation. This transformation results in the following effective $n-1$ spin Hamiltonian. 
\begin{equation}
H'_{SK}=-\sum_{i<j}^{n-1}J_{ij}\sigma_i^z \sigma_j^z+\sum^{n-1}_{i=1} h_i \sigma^z_i, \label{eq:SK_fix}
\end{equation}
where $h_i=J_{in}$. For the proof-of-principle I generate $1500$ such Hamiltonians with $n=17$. I then run Path Integral Quantum Annealing (PIQA) $1001$ times for each such Hamiltonian, following the methods used in  \cite{Chancellor2016}, which were adapted from those in \cite{Martonak2002}, but with $T=0.8246$, $\tau=20$ and $P=30$. For each spin within each Hamiltonian, I compare the average value of the annealer output to a simple certainty value $P_i$ calculated using 
\begin{align}
S_i=\mathrm{sgn}( \sum_{j=1}^{N} G_j), \label{eq:S_raw} \\
P_i=\frac{ \sum_{j=1}^{N} \delta_{G_j,-S_i}}{N}, \label{eq:P_raw}
\end{align}
where $G$ consists of the list of the $1001$ solutions returned by PIQA ($G_i \in \{1,-1\}$). I then break these spins up into two categories, those where $S_i$ found by Eq.~(\ref{eq:S_raw}) agrees with the true solution found by exhaustive classical search, and those where it does not. As Fig.~\ref{fig:counts_P} clearly shows, the larger the value of $P_i$ becomes, the more likely it is that the bit value is incorrect. Therefore the statistics of our simulated quantum annealer outputs not only information about the probable value of a bit in a given solution, but also about the relative certainty of different bit values. How effectively this information is used depends on the heuristic used in $\mathcal{F}$, I discuss a few examples of how $\mathcal{F}$ could be constructed in Sec.~\ref{F_heurists}.

\subsection{Definitions}
\begin{figure}
\begin{centering}
\includegraphics[width=7cm]{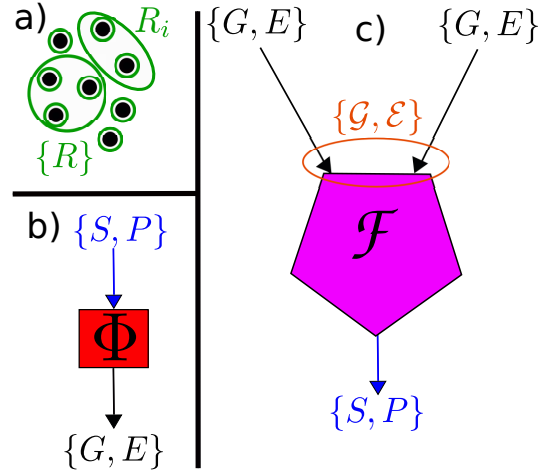}
\par\end{centering}
\caption{\label{fig:visual_defs} Visual explanation of functions used within an inference primitive protocol.  a) Sets of one or more bits (black circles) $\{R\}$ represented by green ovals, b) inference primitive $\Phi$, c) processing function $\mathcal{F}$. All quantities are defined in table \ref{tab:definitions}.}
\end{figure}
I now define a mathematical representation of the computational subroutine I have described earlier. Firstly I consider a system of $N_{bits}$ bits. To simplify some mathematical definitions which I will give later and for consistency with spin Hamiltonian definitions, I allow these bits to take values $\{1,-1\}$, rather than $\{1,0\}$. I further define clusters $R_i$ which each consist of a unique, non-empty  set of these bits, as represented in Fig. \ref{fig:visual_defs}(a). 

I also define an \emph{inference primitive} $\Phi$, which takes as inputs a list of guesses for the value of the bits, $S$, as well as uncertainty values $P$ for each cluster in $R$.  An inference primitive in turn outputs a list of solution candidates $G$, and a list of associated energies for each candidate $E$. Each solution candidate consists of  $N_{bits}$ numbers, each corresponding to a bit value of $\{1,-1\}$.  The energy value $E_i=\sandwich{G_i}{H_{problem}}{G_i}$ tells how optimal each solution value is, where lower values indicate a higher level of optimality. Lists $G$ and $E$ must have the same length, which I refer to as $N_{out}$.  Fig. \ref{fig:visual_defs}(b) represents an inference primitive visually. In practice, the role of $\Phi$ will be played by a call to an analog computational element, in the case of this paper, a quantum annealer.

In the absence of multi-bit clusters, $S$ and $P$ could be defined as a single `mean' bit value for each bit which could be written as $v_i=(1-P_i)\,S_i\in [-1,1]$. However, this notation does not easily generalize to include multi-bit clusters, and therefore I represent $S$ and $P$ as distinct quantities where $|S|\le|P|$. Parametrizing in terms of $S$ and $P$ is natural as these two quantities map to different control parameters within an annealing protocol.

In addition to the inference primitive, I also define a mathematical function which I call the \emph{processing function} $\mathcal{F}$. This function takes as its input a list of lists $\mathcal{G}$, each element of which is a list $G$ of solution candidates. This function likewise takes $\mathcal{E}$ as an input, which is a list of lists $E$ of the associated energies for each solution candidate. The lists $\mathcal{G}$ and $\mathcal{E}$ must have the same length which I call $N_{inputs}$.  Generally,  $\mathcal{G}$ and $\mathcal{E}$  will be allowed to be empty ($N_{inputs}=0$). This function outputs a list of guesses for the values of each of the bits $S$, and an uncertainty value $P$ for each cluster in $R$. A processing function is represented visually in Fig. \ref{fig:visual_defs}(c).

I have now defined an inference primitive $\Phi$, the outputs of which can be used to construct the inputs of a processing function $\mathcal{F}$, in turn the outputs of $\mathcal{F}$ can be used as the inputs of $\Phi$. The mathematical functions and their associated inputs and outputs define the basic structure of the inference primitive framework, these mathematical functions can be expressed diagrammatically as depicted in Fig.~\ref{fig:visual_defs} and this diagrammatic representation can be used to express sophisticated protocols as discussed in Sec.~\ref{sub:exist_examples} and \ref{sub:alg_struct}.

It is useful to give a few more definitions of mathematical quantities which will become important in specific examples which I will give later in this paper. In particular, to define ways in which $\mathcal{G}$ and $\mathcal{E}$ can be reduced to lists, rather than lists of lists. I first consider `flattened' versions of the lists $\mathcal{G}$ and $\mathcal{E}$, $\tilde{G}=\mathcal{G}_1\cup \mathcal{G}_2 \cup...$ and $\tilde{E}=\mathcal{E}_1\cup \mathcal{E}_2 \cup...$, both will have length $N_{flat}=N_{inputs}\,N_{out}$. These flattened versions contain all of the information within the original lists $\mathcal{G}$ and $\mathcal{E}$ except for information about where each solution candidate came from. As I will discuss later, many processing functions may be constructed for which information about where each solution candidate originated is not important. A second pair of useful quantities is the list of \emph{unique} solution candidates in  $\tilde{G}$, and their associated energies. I label these quantities  $\tilde{G}^{(u)}$ and  $\tilde{E}^{(u)}$, with a new length $N_u\le N_{flat}$.

As a convention, for $\tilde{G}$ and $\tilde{G}^{(u)}$, which are both solution candidate lists, I use a subscript to refer to the solution number and put the list of bits to be considered as a functional argument. For instance $\tilde{G}_j(i)$ is the value of the $i$th bit in solution candidate number $j$. Alternatively, $\tilde{G}_j[R_i]$ is the list of all of the bit values over the cluster $R_i$ in solution candidate number $j$. For $S$, which only has a bit index, I use the subscript to refer to the bit cluster, so for instance $S_i$ refers to the value of the inferred bit value of bit $i$ and while $S_{R_i}$ refers to the list of inferred bit values on the cluster $R_i$, expressed mathematically $S_{R_i}=\{S_x: x \in R_i\}$.

For single bit clusters, the solution candidates can be divided into two groups based on the value of the bit. For multi-bit clusters the picture is more complicated, one quantity which I will demonstrate later is convenient to define is a \emph{weighting factor}, $W(\tilde{E}_j,\tilde{G}_j[R_i],S_j)$ which weights the importance of each state to calculating $P$ for the cluster. Based on these weighting factors, I define
\begin{equation}
P_i=\mathrm{min}\left(\frac{\sum_{M_j<0}W(\tilde{E}_j,\tilde{G}_j[R_i],S_{R_i}) }{\sum_{\forall j} W(\tilde{E}_j,\tilde{G}_j[R_i],S_{R_i})},0.5\right)\label{eq:P_Wfactor},
\end{equation}
where $M_j=\sum_{k\in R_j} S_k \frac{\tilde{G}_j(k)}{|R_j|}$, and the minimum value is taken to guarantee that $P_i \in [0,0.5]$.  For simplicity,  one can further restrict this study to functions $W$ which can be decomposed into two parts, one which depends purely on $\tilde{E}$, and one which depends purely on $S$ such that 
\begin{equation}
 W(\tilde{E}_j,\tilde{G}_j[R_i],S_{R_i})= \hat{W}(\tilde{E}_i)\bar{W}(\tilde{G}_j[R_i],S_{R_i}).\label{eq:W_decomp}
\end{equation}
As a further matter of notation, I use piping symbols $|\star|$ to refer to the length of a list, so for instance $|R|$ means the number of elements in the list $R$.

\begin{table*}
\begin{centering}
\begin{tabular}{|c|c|c|}
\hline 
 Quantity &  Definition & Properties \tabularnewline
\hline 
\hline 
$R$ & Set of list of bits involved in each cluster & $R_i =\{m : m \in \mathbb{Z}_{N_{bits}}\}$, $|R|\ge N_{bits}$ \tabularnewline
\hline 
 $S$ & Inferred value for each bit & $S_i\in \{-1,1\}$, $|S|=N_{bits},S_{(R_i)}=\{S_m: m \in R_i\}$ \tabularnewline
\hline 
 $P$ & Uncertainty in the value of each cluster of bits $m_i$ & $P_i\in [0,0.5]$, $|P|=|R|$ \tabularnewline
\hline 
 $G$ & List of solution candidates & $ G_j=\{ q: \, \{q_j \in \{ -1, 1 \}\}, |q|=N_{bits} \}$, $|G|=N_{out}$   \tabularnewline
\hline 
 $E$ & Solution candidate energies & $E_j=\sandwich{G_j}{H_{problem}}{G_j}$, $|E|=N_{out}$   \tabularnewline
\hline 
$\mathcal{G}$ & set of different $G$ & $ \mathcal{G}_k= G$, $|\mathcal{G}|=N_{inputs}$   \tabularnewline
\hline 
$\mathcal{E}$ & set of different $E$ & $ \mathcal{E}_k=E$, $|\mathcal{E}|=N_{inputs}$   \tabularnewline
\hline 
$\tilde{G}$ & $\tilde{G}=\bigcup_r \mathcal{G}_r=\mathcal{G}_1\cup \mathcal{G}_2 \cup...$ & $|\tilde{G}|=N_{flat}=N_{inputs}\, N_{out}$,   \tabularnewline
\hline 
$\tilde{E}$ & $\tilde{E}=\bigcup_r \mathcal{E}_r=\mathcal{E}_1\cup \mathcal{E}_2 \cup...$ &  $|\tilde{E}|=N_{flat}=N_{inputs}\, N_{out}$   \tabularnewline
\hline 
 $\tilde{G}^{(u)}$ & List of all unique solution candidates in  $\tilde{G}$& $ \tilde{G}^{(u)}_i=\{ q:\, \{q_j \in \{ -1, 1 \}\}, |q|=N_{bits} \}$, $|G^{(u)}|=N_{u}$   \tabularnewline
\hline 
 $\tilde{E}^{(u)}$ & Unique solution candidate energies &  $|\tilde{E}^{(u)}|=N_{u}$   \tabularnewline
\hline 
 $\mathcal{F}$ & Map from $\mathcal{G}$ and $\mathcal{E}$ to $P$ and $S$ given $R$ &  $\mathcal{F}:\{ \mathcal{G},\mathcal{E},R\} \mapsto \{P,S\}$   
\tabularnewline
\hline
$\Phi$ & Inference primitive & $ \Phi: \{P,S,R\} \mapsto \{G,E\}$  
\tabularnewline
\hline
$W$ & Weighting factor sometimes used to calculate $P$ & Eq. \ref{eq:P_Wfactor}, Eq. \ref{eq:W_decomp}
\tabularnewline
\hline 
$\hat{W}$ & Energy dependent part of weighting factor $W$ &  Eq. \ref{eq:W_decomp}
\tabularnewline
\hline 
$\bar{W}$ & Bit value dependent part of $W$ &  Eq. \ref{eq:W_decomp}
\tabularnewline
\hline 
$\tilde{G}_k(l)$ & Notational shorthand used with $\tilde{G}$ and   $\tilde{G}^{(u)}$  &  $\tilde{G}_k(l)=\{x_l : x=\tilde{\mathcal{G}}_k\}$
\tabularnewline
\hline 
$\tilde{G}_j[R_i]$ & Notational shorthand used with $\tilde{G}$ and   $\tilde{G}^{(u)}$ &  $\tilde{G}_k[R_i]=\{\tilde{G}_j(y): y \in R_i \}$
\tabularnewline
\hline 
\end{tabular}
\par
\end{centering}
\caption{List of quantities and their definitions, I use piping symbols $|\star|$ to refer to the length of a list, so for instance $|R|$ means the number of elements in the list $R$. \label{tab:definitions}}

\end{table*}

\subsection{Examples with Existing protocols \label{sub:exist_examples}}

Let us now discuss in more detail how to construct algorithms based on inference primitives from quantum annealers. As an example, I will first explicitly demonstrate how both the traditional QAA and the simplest local search method of \cite{Chancellor2016} can be re-expressed in terms of inference primitives. 

The traditionally formulated QAA is not biased toward a particular state, we formulate a processing function $\mathcal{F}_{init}$ which takes no inputs and returns $P_i=0.5\,\forall i$. For these values of $P$, the values of $S$ do not matter, so we set them to be all $1$ without loss of generality,
\begin{equation}
\mathcal{F}_{init}:\{\{\},\{\},R\}\mapsto \{\{0.5,0.5,...\},\{1,1...\}\}
\label{eq:init_QAA}
\end{equation}

\begin{figure}
\begin{centering}
\includegraphics[width=4cm]{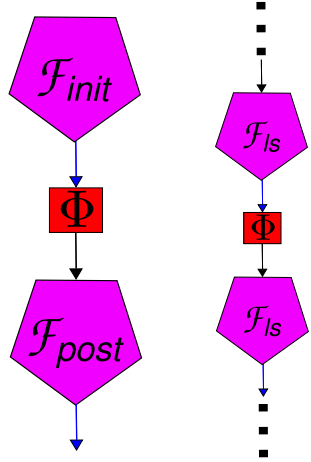}
\par\end{centering}
\caption{\label{fig:old_prots}Left: Traditional QAA formulated in terms of inference primitives and processing functions, where $\mathcal{F}_{init}$ is defined in Eq.~(\ref{eq:init_QAA}). Right: local search protocol formulated in terms of inference primitives and processing functions, the general for $\mathcal{F}_{ls}$ is given in Eq.~(\ref{eq:F_ls}).}
\end{figure}

In general, the traditional QAA can be augmented by sophisticated post processing, \cite{Nishimura2016, qbsolve,Bian2014,Bian2016}, and therefore after the inference primitive, we should include a second processing function $\mathcal{F}_{post}(G,E,R)$ to include all of these possibilities. This representation is depicted on the left of Fig.~\ref{fig:old_prots}. The hybrid methods used in \cite{qbsolve,Bian2014,Bian2016} actually use multiple runs with changing problem definitions to solve a problem, and therefore constitute many repeated runs of the protocol depicted on the left of Fig.~\ref{fig:old_prots}. I discuss in Sec.~\ref{sub:belief_propagation} how such existing hybrid techniques may be combined with more sophisticated inference primitive protocols.

 For the local search protocols considered in \cite{Chancellor2016}, the results of previous calls to the inference primitive are used sequentially, with the result of a previous call being fed into the next iteration of the protocol, as depicted on the right of Fig.~\ref{fig:old_prots}. In this case, however, there is only one global value of $P_i=p\,\forall i$ which defines the uncertainty, the processing function which is run at each step can therefore be defined as 
\begin{align}
\Phi:\{\{p,p,...\},S,R\} \mapsto \{G,E\}, \nonumber \\ 
\mathcal{F}_{ls}:\{G,E,R\}\mapsto\{\{p',p',...\},S'\},
\label{eq:F_ls}
\end{align}
where $p'$ is the global value of $P$ to be used for the next local search, and the protocol is run iteratively with $p\leftarrow p'$ and $S\leftarrow S'$ at each step. This formalsim can further be generalized to represent another class of hybrid annealer based algorithms, which can be used without any reverse annealing capabilities. These algorithms, which have been shown to be successful in \cite{Karimi2017,Karimi2017a} work by `fixing' some qubits by removing them from the problem description and replacing them with appropriate field terms to match the state which they are assumed to take. This kind of process allows an annealer without reverse annealing to be represented by an inferrence primitive where $p_i$ is restricted to only take values of either $0$, indicating that a spin is to be `fixed' or  $0.5$ for those which are not removed and will be annealed normally. The representation of these algorithms in the inferrence primitive formalism are therefore exactly the same as the ones for the local search given in Fig.~\ref{fig:old_prots}, but with
\begin{align}
\Phi:\{P,S,R\} \mapsto \{G,E\}, \nonumber \\ 
\mathcal{F}_{fix}:\{G,E,R\}\mapsto\{P',S'\},
\label{eq:F_fix}
\end{align}
where $P'_i \in \{0,0.5\}$.

\begin{figure}
\begin{centering}
\includegraphics[width=4cm]{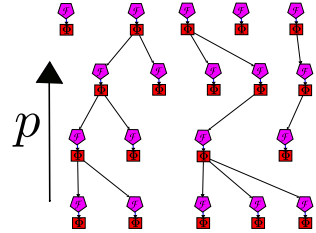}
\par\end{centering}
\caption{\label{fig:pop_anneal_noGen}Structure of the poplution annealing inspired protocols from \cite{Chancellor2016} expressed in the inference primitive formalism.}
\end{figure}

\begin{figure}
\begin{centering}
\includegraphics[width=4cm]{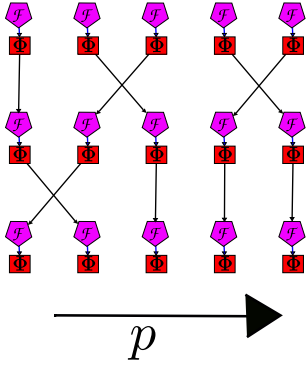}
\par\end{centering}
\caption{\label{fig:par_temp_noGen}Structure of the parallel tempering inspired protocols from \cite{Chancellor2016} expressed in the inference primitive formalism.}
\end{figure}

Going beyond simple local search \cite{Chancellor2016}, protocols incorporating local search that are inspired by the state-of-the-art optimization techniques of parallel tempering \cite{Swendsen1986, Earl2005} and population annealing  \cite{Hukushima2003,Matcha2010,Wang2015,Barzegar2017}, these algorithms can be represented in this framework. The processing function and inference primitives will still have the general local search structure in Eq.~\ref{eq:F_ls}, but generally allow $\{G,E\}$ to be copied (in the case of population annealing) or exchanged between sets of inferrence primitives with different $p$ values. The structure of the population annealing inspired protocol is depicted in Fig.~\ref{fig:pop_anneal_noGen}, while the structure of a parallel tempering inspired algorithm is depicted in Fig.~\ref{fig:par_temp_noGen}.

\section{Algorithmic Design\label{alg_design}}

As well as being a powerful tool for expressing currently proposed algorithms, the inference primitive formalism is also a powerful tool in designing new algorithms. This formalism depicts the different possible ways for information to flow between classical processing and a quantum `inference primitive' subroutine in a high level way, and therefore can be used to express different algorithmic possibilities in terms of information flow. Thus far, we have only considered processing functions which take outputs from a single call to an inference primitive, however, processing functions can be constructed which take information from multiple inference primitive calls. Using processing functions in this way represents a breeding hybridization step in a genetic algorithm. While the focus of this paper is on developing the inference primitive formalism for design of annealer algorithms, rather than to design specific heuristics, it is still useful to discuss examples how different processing function heuristics can be constructed, which I do in the next subsection.

\subsection{Processing Function Heuristics \label{F_heurists}}

Although the primary purpose of this paper is not to design algorithms, it is worth briefly discussing what form the heuristics in the processing function could take, including some examples which are direct extensions of work which has already been done. While testing these heuristics would be useful, doing it properly would be quite an involved task, and therefore beyond the scope of the current work. The focus of this work is to examine how new algorithms can be designed for a quantum annealer with generalized controls, not to study relative algorithm performance. 

Recall that I have already discussed heuristics to convert probability values for each qubit into the actual $s'$ values which will be supplied to the annealer. In the inference primitive formalism details of the exact experimental implementation are contained with the inference primitive $\Phi$ itself, rather than the processing function $\mathcal{F}$. In this subsection however, I focus on the processing function $\mathcal{F}$ which provides uncertainty information which can then be converted to experimental parameters in the inference primitive.  

For simplicity, let us start with cases where the processing function $\mathcal{F}$ only has a single stream of input values from the inference primitive $\{G,E\}$. In this case, the simplest thing to do is just to take statistics over the raw data, calculating the probability that a bit will take a certain value directly by averaging over $G$ with no regard for $E$, as was done in Eq. \ref{eq:S_raw} and \ref{eq:P_raw}. Such a simplistic approach relies on the ability of the inference primitive, for instance a quantum annealer, to always find highly optimal states. However, in practice real devices may not do this. 

One approach to mitigate the fact that some solutions in $G$ may not be very optimal is to only consider candidates which have an energy below an `elite threshold', this approach has already proved useful in hybrid algorithms used in \cite{Karimi2017,Karimi2017a} which do not require an initial state to be seeded. Those papers, however, were based on annealers which did not have reverse annealing capabilities. With reverse annealing capabilities (and independent annealing controls of individual qubits), their method can be extended to include the possibility where the direction of a state of a qubit is suspected but should not be assigned with $100\%$ certainty. A simple generalized processing function in this case could take the form:
\begin{align}
S_i=\mathrm{sgn}(\sum_{j=1}^{N} G_j)  \Theta(E_{elite}-\tilde{E}_j)), \label{eq:S_elite} \\
P_i=
\frac{\mathrm{min}( \sum_{j=1}^{N} \delta_{G_j=S_i}\Theta(E_{elite}-E_j))}{\mathrm{min}( \sum_{j=1}^{N}\Theta(E_{elite}-E_j))}, \label{eq:P_elite}
\end{align}
where $\Theta$ is the Heaviside step function defined so that  $\Theta(a)=1$ if $a>0$ and $\Theta(a)=0$ otherwise, and $E_{elite}$ is the elite energy threshold, as assigned in \cite{Karimi2017,Karimi2017a}. Note that, as was previously done in this algorithm, any qubit with $P_i=0$ can be excluded from the actual annealer run and replaced with field terms.

Rather than using a hard cutoff, another way to give preference to low energy solution candidates when calculating $S$ and $P$ is to thermally reweight each of the unique candidates
\begin{align}
S_i=\mathrm{sgn}(\sum_{j=1}^{N_{u}} G^{(u)}_j \exp(-\frac{E^{(u)}_j}{T})), \label{eq:S_therm} \\
P_i=\frac{1}{Z}(\sum_{j=1}^{N_u} \delta_{G^{(u)}_j,-S_i} \exp(-\frac{E^{(u)}_j}{T})), \label{eq:P_therm}
\end{align}
where the $(u)$ superscript indicates a set of solution candidates and energies where duplicate candidates in $G_i$ have been removed. In this case, $T$ can be thought of as a meta-parameter which controls the effective range of the search that will be performed by the inference primitive. This suggests that one algorithmic possibility could be to run a series of inference primitive calls as depicted in Fig.~\ref{fig:old_prots}(right), but with successively decreasing $T$ as a simulated annealing analogue.

Thus far we have only considered processing functions $\mathcal{F}$ which take a single $\{G,E\}$, however, for genetic algorithms, we need to define processing functions which take sets of inference primitive outputs $\{\mathcal{G},\mathcal{E}\}$. One way to construct such processing function heuristics is to create flattened lists, which treat all solution candidates as if they came from a single inference primitive, these flattened data $\{\tilde{G},\tilde{E}\}$ can then be used directly in heuristics such as those discussed earlier in this section. Not all processing functions can be represented in this way, however, for example a processing function $\mathcal{F}$ could take the lowest energy solution candidate from two different $\mathcal{G_i}\in\mathcal{G}$ and assign $P_i=0.5$ to bits which disagree between the two and $P_i=p$ where $0<p<0.5$ to those which do.
 
\subsection{Algorithm Structure\label{sub:alg_struct}}
\begin{figure}
\begin{centering}
\includegraphics[width=4cm]{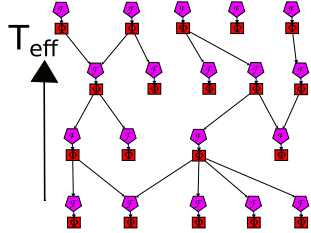}
\par\end{centering}
\caption{\label{fig:pop_anneal_gen}Structure of population annealing inspired protocols with additional genetic component.}
\end{figure}

\begin{figure}
\begin{centering}
\includegraphics[width=4cm]{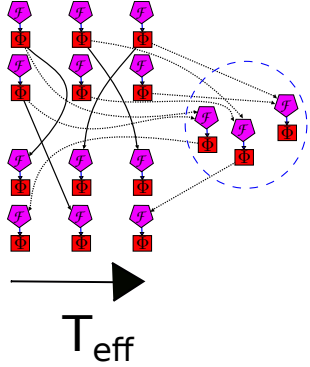}
\par\end{centering}
\caption{\label{fig:par_temp_gen}Structure of parallel tempering inspired protocols with additional genetic component. The `genetic pool' of inference primitives and processing functions is circled in blue dashed lines.}
\end{figure}

Now that I have given examples of how processing function heuristics can be constructed, it is worth briefly considering how the inference primitive formalism can be used as a graphical tool to design new algorithms. For instance, a genetic component can be added to the population annealing algorithm depicted in Fig.~\ref{fig:pop_anneal_noGen} by allowing multiple edges to be incident on each processing function, as depicted in Fig.~\ref{fig:pop_anneal_gen}. Because of the way the total population is controlled in these algorithms (see \cite{Hukushima2003}), adding a fixed number of extra processing functions which accept two or more inputs to produce offspring will not cause the population to grow (or shrink) uncontrollably. In this example, which inference primitive outputs get to produce extra offspring could be chosen for instance by drawing two or more from a Boltzmann probability distribution constructed from the lowest energy given by each inference primitive call (as was suggested in \cite{Chancellor2016}) $P_j=\exp(-\min (\mathcal{E}_j)/T_{eff})/Z$ without replacement.
The inference primitive formalism can also demonstrate how we can add a genetic component to a parallel tempering inspired algorithm. In such an algorithm one can replace each single call to an inference primitive at an effective temperature with a pair of calls, and than combine these outputs in a `hybridization pool' consisting of inference primitive calls based on pairs of inference primitive outputs as depicted in Fig.~\ref{fig:par_temp_gen}. These hybridization results could then be reinserted into the main pool of inference primitive calls probabilistically, one way to accomplish this is to use the process outlined below:
\begin{enumerate}
\item Produce `genetic pool' of inference primitive outputs, for instance using some of the methods discussed in the previous subsection.
\item For each set of inference primitive outputs in the genetic pool, $\{G^{hyb},E^{hyb}\}$, starting from the lowest $T_{eff}$ and increasing, have this set of outputs replace a set in the standard inference primitive pool probabilistically with a probability determined by
\begin{equation}
P_{ex}=\min\left(\exp(\frac{\min (E^{hyb})-\min (E)}{T_{eff}}),1\right), \nonumber
\end{equation}
where $T_{eff}$ is the effective temperature which has been used on the inference primitive in the parallel tempering pool. If either a replacement has been performed, or all inference primitive outputs in the regular parallel tempering pool have been tested and none have been replaced, move on to the next set of hybridzation outputs. In the case where a replacement has been successfully performed discard the inference primitive outputs which have been replaced, otherwise, discard the outputs in the genetic pool. Once all outputs in the gentic pool have been either discarded or used as replacements, move on to the next step. 
\item Perform parallel tempering inspired swaps using the standard update rules as described in \cite{Chancellor2016}.
\end{enumerate}

There are also many other algorithms which can be discovered using the inference primitive formalism. The two ideas here are included to give examples of how the inference primitive formalism can be used as a tool to visualize information flow in annealer based algorithm design.

\section{Compatibility with Other Methods\label{method_comp}}

Now that I have demonstrated the power of the inference primitive formalism in terms of designing algorithms based on quantum annealers with generalized classical controls, I turn my attention to how these methods are compatible with many methods which currently represent the state of the art, as well as techniques which are now on the horizon. This section is not supposed to be an exhaustive list, but rather to give the reader an idea of the versatility of inference primitive based annealer computation. 

\subsection{Protocol Modifications \label{sub:protocol}}

\begin{figure}
\begin{centering}
\includegraphics[width=7cm]{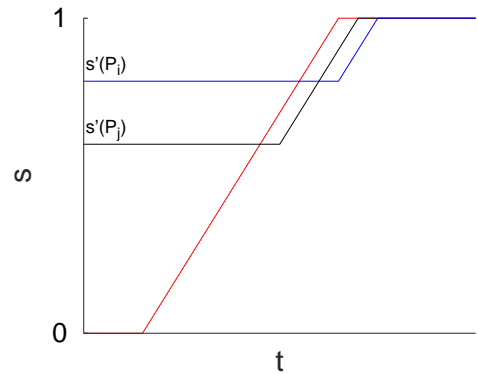}
\par\end{centering}
\caption{\label{fig:freeze_sync} Depiction of how the time at which annealing from $s'$ is started can be used to advance or retard individual qubit annealing schedules to synchronize freezing.}
\end{figure}
The first technique which I discuss are techniques developed by D-Wave Systems Inc. to advance or retard individual qubits to synchronize freezing \cite{LantingAQC2016} using an effective local temperature estimated using the methods in \cite{Raymond2016}. These methods apply to the relative values of the annealing parameter $s$ during the final forward anneal, a parameter which is not fixed by the inference primitive protocol described in Sec.~\ref{sec:inf_prim}, and therefore freezing can be synchronized by advancing or retarding the point at which one qubit begins its final forward anneal relative to the other qubits, as depicted in Fig.~\ref{fig:freeze_sync}.
\subsection{Higher Order Drivers\label{sub:high_order}}
Let us now consider generalizations of inference primitive protocols for multi-body drivers, which are necessary to realize non-stoquastic drivers, for example. Previously, $R$ has just been a list of every qubit, but now will also include some clusters of qubits which are flipped simultaneously by multibody drivers. To determine the strength at which multibody drivers are applied, one should consider statistics over the overlap of each of the members of  $G_j$ with the solution candidate $S$ over the relevant cluster, $M_j=\sum_{k\in R_j} S_k \frac{\tilde{G}_j(k)}{|R_j|}$ where $|R_j|$ is the number of elements in $R_j$. When  $M_j=1$, then the cluster agrees exactly for the candidate solution and the $\tilde{G}_j[R_i]$. The value $M_j=-1$ indicates perfect disagreement.  The uncertainty value $P_i$ for the cluster $R_i$ corresponds to the probability that $S_{R_j}$ is  closer in Hamming distance to the correct solution than $\neg S_{R_j}$. Positive $M_j$ indicates that  $S_{R_j}$  is the closer of the two, whereas negative indicates that $\neg S_{R_j} $ is closer. 

For each cluster, we formulate a weighted sum to determine the probability that $S_{(R_i)}$ is closer. To achieve this, I define $P$ in terms of a weighting factor $W$ using Eq. ~(\ref{eq:P_Wfactor}). For simplicity,  I assume that $W$ can be decomposed into two terms such that $ W(\tilde{E}_j,\tilde{G}_j[R_i],S_{(R_i)})= \hat{W}(\tilde{E}_i)\bar{W}(\tilde{G}_j(R_i),S_{(R_i)})$.  For the energy dependent part, one could for example  define $\hat{W}(\tilde{E}_i)=\exp(\frac{-\tilde{E}_i}{T})$ corresponding to the thermal weighting as in Eq.~(\ref{eq:P_therm}), $\hat{W}(\tilde{E}_i)=1$ for unweighted averages as in Eq. (\ref{eq:P_raw}), or finally $\hat{W}(\tilde{E}_i)=\Theta(\tilde{E}_{elite}-E_i)$ for a multi-bit analogue of  the elite averages used in Eq. (\ref{eq:P_elite}). As for $\bar{W}(\tilde{G}_j[R_i],S_{(R_i)})$, it should be weighted to favor $|M_j|$ close to 1, as these are the values for which cluster flips will make the largest difference. A logical choice is therefore to choose weights which are inversely proportional to the number of states within the same Hamming distance from either $S_{(R_j)} $ or $\neg S_{(R_j)} $, 
 
 \begin{equation}
\bar{W}(\tilde{G}_j[R_i],s_j)=\binom{|R_j|}{\mathcal{D}(S_{(R_i)},\tilde{G}_j[R_i])}^{-1}
 \end{equation}
 where $|R_j|$ indicates the number of elements in the set, and  $\mathcal{D}(S_{(R_i)},\tilde{G}_j[R_i])$ indicates the Hamming distance between the two lists.
\subsection{Belief Propagation \label{sub:belief_propagation}}
For the current generation of annealers, with hardware graphs which are relatively small compared to the size of many relevant problems, it is important to be able to solve problems which are larger than the available hardware graph. The general method to do this is to solve problems on modified subgraphs of the hardware graph in an algorithmic way \cite{Bian2014,Bian2016, qbsolve}, eventually converging on a single consistent solution. In this paper I will focus on one particular method, the generalized belief propagation method proposed in \cite{Bian2016} based on earlier work in \cite{Yedida2005}. Although only exact for tree graphs, belief propagation has proven to be an important tool for solving a host of important real world problems, most notably decoding Low Density Parity Check Codes (LDPC) \cite{Kschischang2006,Mceliece1998}. The belief propagation method described in \cite{Bian2016} performs belief propagation between hardware-sized subgraphs to obtain an approximate thermal distribution.

Because this method obtains a distribution, rather than a single state, it can be used effectively as an inference primitive  and therefore can be used as a subroutine in all of the previously discussed algorithms, using the same $\{R, S,P\}$ throughout the protocol until either convergence is found or a timeout occurs. However, the marginals which are calculated throughout the protocol carry beliefs about the likely value of a bit and its uncertainty. The protocol can be made more efficient by using this information to update $\{S,P\}$, whenever the beliefs are updated. With fixed $\{S,P\}$ new information about bit values is wasted. If one of the bit values $S_i$ with a low value of $P_i$, became inconsistent with the others during the course of this protocol it would likely not be able to correct for this inconsistency and may either fail to converge or return a low quality solution.

In the algorithm proposed in \cite{Bian2016}, each bit has an associated marginal, $b_i(z_i)$, which contains information about the relative likelihood of a bit having a value of $1$ or $-1$. Based on a normalized version of this marginal, we can find an approximate value for $S_i$ and $P_i$ which dynamically updates at each step of the protocol:
\begin{align}
S_i=\mathrm{sgn}(b_i(z_i=+1)-b_i(z_i=-1)),\\
P_i=0.5\,\left(1-\left|\frac{b_i(z_i=+1)-b_i(z_i=-1)}{b_i(z_i=+1)+b_i(z_i=-1)}\right|\right).
\end{align}
\section{Conclusions\label{conclusions}}

In this paper I have proposed a new way of thinking about algorithms based on a quantum annealer with generalized classical controls. I have given examples both of how existing quantum annealer based algorithms can be represented in this formalism and how this formalism can be used to design new algorithms, including algorithms with genetic components.  While the algorithms proposed here will not in general obey detailed balance, they could allow for a more complete accounting of the low energy local minima of an energy landscape, and therefore may be useful for calculating thermal distributions if used with appropriate post processing. To motivate this formalism I have given a proof-of-principle demonstration that the output of annealer runs contain information not only about the likely solution to a problem Hamiltonian, but also the relative bitwise uncertainty.

Although a full analysis is beyond the scope of this paper, it would likely be interesting to explore the connection between the methods proposed here and quantum inspired diffusion Monte Carlo algorithms as discussed in \cite{Jarett2016,Jarett2017}, which show similar structure in the methods with which they solve problems. It would likewise be interesting to develop inference primitives based on other physical mechanisms, such as closed system adiabatic quantum computing, or quantum walks. It would also be interesting to run comparisons of algorithms designed with this formalism on real devices to determine their performance, and to design more algorithms. The algorithms given in this paper are only intended as examples of how the design techniques I have developed can be used, this paper has only scratched the surface of the algorithmic possibilities for this functionality of a quantum annealer.

\section*{Acknowledgments}

The author was supported by EPSRC ( grant ref: EP/L022303/1), and
would like to thank Viv Kendon, Joschka Roffe, and Dominic Horsman for critical reads of the
paper and useful discussions. The author would also like to thank Alejandro Perdomo-Ortiz for making me aware of the alternative protocol in \cite{Perdomo-Ortiz2011}.

\section*{Competing Interests}

I declare that I have no competing interests.

\section*{Appendix 1: Example of a Heuristic to Relate Uncertainty to Transverse Field}

There are many potential heuristics which could be used to relate the probabilites $P$ which are passed to an inference primitive to the annealing $s'$ parameter which is use in a reverse annealing protocol. While the focus of this paper is not on how to actually relate these two parameters, it is instructive to give a simple example of what one such heuristic could look like. Whether or not this heuristic works well in practice is beyond the scope of this current work, and almost certainly more sophisticated heuristics, for instance based on the local temperture estimates given in \cite{Raymond2016} are likely to perform better.

To start with, I make use of the fact that it has been numerically demonstrated that quantum fluctuations moderated by a transverse field can be used as a proxy for thermal fluctuations for inference problems \cite{Otsubo2012}. In this spirit I define an approximate effective temperature related to a transverse field strength, which is set by a chosen value of $s$ in Eq.~(\ref{eq:H_Anneal})  which I denote as $s'$. This can be done using the method suggested in \cite{Chancellor2016} by analytically diagonalizing the Hamiltonian at the appropriate point in the annealing schedule with a "problem" Hamiltonian consisting of a single bit Hamiltonian with a longitudinal field of unit strength, $H_{1}(s')=-A(s')\,\sigma^{x}+B(s')\,\sigma^{z}$.  This ratio is then compared to a Boltzmann distribution, and  the equation inverted to solve for temperature. This approach yields
\[
T'(s')=
\]
\begin{equation}
2\,\left[\ln\left(\left|\frac{\sqrt{A(s')^{2}+B(s')^{2}}}{A(s')}+\frac{B(s')}{A(s')}\right|^{2}\right)\right]^{-1}.\label{eq:Teff}
\end{equation}

In situations where coupling is present, rather than the single qubit case examined here, the effective picture becomes more complicated. To correctly determine the effect of a coupler on a single qubit, one must take into account the fact that all other qubits within the coupler are also fluctuating in a way which is generally complicated and correlated both with each other and the qubit we are examining. The results in \cite{Otsubo2012} suggest, however, that these complicated effects will be very similar for both quantum and thermal fluctuations. Based on these similarities, a simple first approximation is to apply relationships between temperature and driver strength which are derived in the single qubit case to larger multi-qubit systems, based on the reasoning that the effects of correlations with neighbors may be qualitatively similar in both cases. While this is a rather crude approximation, the heuristic given here is only intended as a minimal example, single qubit dynamics provide one of the simplest ways to relate temperature to transverse field. Alternatively, a local temperature could be estimated experimentally  using the methods of \cite{Raymond2016}, or by adapting the methods to estimate a global effective temperature used in \cite{Benedetti2016}.

Now I use the seminal result by Nishimori \cite{Nishimori1980,Nishimori2001,Nishimura2016}  that a temperature can be related to an error probability via the Nishimori temperature, $T_N$. This relationship is mathematically rigorous and is the underlying principle behind maximum entropy inference, which has many practical applications \cite{Frieden1972,Berger1996,Phillips2006,Raychaudhur2002,Gilmore1996,Mistrulli2011,Rujan1993}.  In these applications, the Nishimori temperature 
\begin{equation}
T_N=2\,\left[\ln(\frac{1-P}{P})\right]^{-1},
\end{equation}
 serves to match a temperature to an effective uncertainty, expressed as a probability $P$. The quantity could be, for instance, an error rate in the context of decoding of communications as in \cite{Rujan1993}. In the context of inference primitive protocols, $P$ should be taken as $P_i$ for a given bit or cluster of bits  a simple approximate heuristic to relate the probailities to the effective temperature $T'$ is to set it to be proportional to the Nishimori temperature 
\[
T'(s')\propto T_N.
\]
By plugging in the approximate formula in Eq.~\ref{eq:Teff} and inverting the equation, I obtain the approximate uncertainty value,
\begin{equation}
P(s')=\left[1+\left|\frac{\sqrt{A(s')^{2}+B(s')^{2}}}{A(s')}+\frac{B(s')}{A(s')}\right|^{2}\right]^{-1}.\label{P_sprime}
\end{equation}
The relationship I have just derived allows a direct definition of the uncertainty values defined in $\{P\}$ in Sec. \ref{sec:inf_prim}  in terms of real device parameters. Expressed in these term, the algorithms in \cite{Chancellor2016} assign the same probability of being incorrect to every bit value. 

Thus far, I have assumed that the annealer is exposed to a bath with a temperature which is low compared with the relevant energy scales $A(s')$ and $B(s')$. However, this may not be the case in a real annealer. In this case we can make the approximation that the themal and quantum fluctuations act in  a statistically independent way and add them in quadrature,
\begin{equation}
T_N=\sqrt{T'^2(s')+\left( \frac{T_{phys}}{B(s')} \right) ^2},
\end{equation}  
where $T_{phys}$ is the physical temperature. Carrying this result through, we arrive at,
\begin{equation}
P(s')=\left[1+\exp  \left( \frac{2}{\sqrt{T'^2(s')+( \frac{T_{phys}}{B(s')} ) ^2}} \right) \right]^{-1}. \label{P_sprimeT}
\end{equation}

It is worth discussing briefly a special subclass of problem Hamiltonians for which $h_i=0 \forall i$ in Eq. (\ref{eq:ISGham}).  For the quantum annealing algorithm applied to such a problem Hamiltoninan, the mean orientation of a bit is zero $\left< \sigma^z_i \right>=0$ and similarly for any cluster of bits $\left<\sum_{j\in R _i} \sigma^z_j\right>=0$  by the fact that these Hamiltonians have a  $\mathbb{Z}_2$ symmetry with respect to flipping all of the qubits (global bit inversion). However, the candidate solution breaks this symmetry, meaning that solution refinement will still work. If multiple sets of annealer outputs are being combined (i.e. $|\mathcal{G}|>1$) for such a problem Hamiltonian, then we should consider the possibility of performing global spin inversions on some of the sets of outputs before applying the processing function. Ideally this should be chosen as the one which yields the highest possible bitwise correlation between all of the candidates. 

Because the space of possible global spin inversions of candidate solutions will be $2^{N_{inputs}}$, performing an exhaustive search over all possible inversions may not be possible if $N_{inputs}$ is moderately large. However a heuristic search method such a simulated annealing could be used to find choices which yield high correlations. Alternatively, one could break the spin inversion symmetry by taking a `majority vote', and performing a global bit inversion on all solution candidates in $\mathcal{G}_k$ if more bits are in the $-1$ state than the $1$ state.

A simple alternative approach for problems where $h_i=0 \forall i$ is to effectively fix a single spin arbitrarily, and replace coupling to that spin with fields. While mathematically correct, this approach has the disadvantage that it gives one spin a `privileged' role in that quantum fluctuations damp out the effect of couplers much more strongly then they do fields because the effect of a coupler is moderated by the fluctuations of two qubits, while the effect of a field is moderated only by the fluctuations of the single qubit it is coupled to.

The methods which I have derived in this section to relate  the local annealing parameter on the real device $s'$ to the uncertainty value $P_i$ are not necessarily unique, there will be other suitable mathematical ways to relate these quantities. For real applications the preferred method may actually be to try different heuristics until one is found which works well, or to try to work out this relationship directly experimentally, for instance by adapting the bisection methods used to find the range of local searches proposed in \cite{Chancellor2016}.


\begin{thebibliography}{10}

\bibitem{Finella1994}
A.~B. Finilla, M.~A. Gomez, C.~Sebenik, and D.~J. Doll.
\newblock Quantum annealing: a new method for minimizing multidimensional
  functions.
\newblock {\em Chem Phys Lett.}, 219:343--348, 1994.

\bibitem{kadowaki1998}
T.~Kadowaki and H.~Nishimori.
\newblock Quantum annealing in the transverse {I}sing model.
\newblock {\em Phys. Rev. E}, 58:5355, 1998.

\bibitem{Kaminsky2002}
W.~M. Kaminsky and S.~Lloyd.
\newblock Scalable architecture for adiabatic quantum computing of {NP}-hard
  problems.
\newblock arXiv:quant-ph/0211152, 2002.

\bibitem{Kaminsky2004}
W.~M. Kaminsky and S.~Lloyd.
\newblock Scalable architecture for adiabatic quantum computing of np-hard
  problems.
\newblock In A.~J. Leggett, B.~Ruggiero, and P.~Silvestrini, editors, {\em
  Quantum Computing and Quantum Bits in Mesoscopic Systems}, pages 229--236.
  Springer US, 2004.

\bibitem{Kaminsky2004a}
W.~M. Kaminsky, S.~Lloyd, and T.~P. Orlando.
\newblock Scalable superconducting architecture for adiabatic quantum
  computation.
\newblock arXiv:quant-ph/0403090, 2004.

\bibitem{Coxson2014}
G.~E. Coxson, C.~R. Hill, and J.~C. Russo.
\newblock Adiabatic quantum computing for finding low-peak-sidelobe codes,
  2014.
\newblock Presented at the 2014 IEEE High Performance Extreme Computing
  conference.

\bibitem{perdomo-ortiz2012}
Alejandro Perdomo-Ortiz, Neil Dickson, Marshall Drew-Brook, Geordie Rose, and
  Alan Aspuru-Guzik.
\newblock Finding low-energy conformations of lattice protein models by quantum
  annealing.
\newblock {\em Scientific Reports}, 2(571), 2012.

\bibitem{Amin2016}
M.~H. Amin, E.~Andriyash, J.~Rolfe, B.~Kulchytskyy, and R.~Melko.
\newblock Quantum {B}oltzmann machine.
\newblock arXiv:quant-ph:1601.02036, 2016.

\bibitem{Benedetti2016}
M.~Benedetti, J.~Realpe-G\'omez, R.~Biswas, and A.~Perdomo-Ortiz.
\newblock Estimation of effective temperatures in quantum annealers for
  sampling applications: A case study with possible applications in deep
  learning.
\newblock {\em Phys. Rev. A}, 94:022308, Aug 2016.

\bibitem{Benedetti2016a}
M.~Benedetti, J.~Realpe-G{\'o}mez, R.~Biswas, and A.~Perdomo-Ortiz.
\newblock Quantum-assisted learning of graphical models with arbitrary pairwise
  connectivity.
\newblock arXiv:1609.02542, 2016.

\bibitem{Adachi2015}
S.~H. Adachi and M.~P. Henderson.
\newblock Application of quantum annealing to training of deep neural networks.
\newblock arXiv:1510.06356, 2015.

\bibitem{Chancellor/Zohren2016}
N.~Chancellor, S.~Zohren, P.~Warburton, S.~Benjamin, and S.~Roberts.
\newblock A direct mapping of max k-{SAT} and high order parity checks to a
  chimera graph.
\newblock {\em Scientific Reports}, 6(37107), 2016.

\bibitem{Marzek}
M.~Marzec.
\newblock {\em Portfolio Optimization: Applications in Quantum Computing},
  pages 73--106.
\newblock John Wiley \& Sons, Inc., 2016.

\bibitem{Chancellor2016}
N.~Chancellor.
\newblock Modernizing quantum annealing using local searches.
\newblock {\em New Journal of Physics}, 19(2):023024, 2017.

\bibitem{D-Wave}
D-wave systems inc. website.
\newblock \url{http://www.dwavesys.com/}.
\newblock Accessed: 2016-08-09.

\bibitem{Bravyi2008}
S.~Bravyi, D.~P. DiVincenzo, R.~I. Oliveira, and B.~M. Terhal.
\newblock The complexity of stoquastic local {H}amiltonian problems.
\newblock {\em Quant. Inf. Comp.}, 8(5):0361--0385, 2008.

\bibitem{Bravyi2009}
S.~Bravyi and B.~M. Tehral.
\newblock Complexity of stoquastic frustration-free {H}amiltonians.
\newblock {\em SIAM J. Comput.}, 39(4):1462, 2009.

\bibitem{Bravyi2014}
S.~Bravyi.
\newblock {M}onte {C}arlo simulation of stoquastic {H}amiltonians.
\newblock arXiv:quant-ph:1402.2295, 2014.

\bibitem{Amin_patent}
M.~H.~S. Amin and W.~M. Johnson.
\newblock Systems and methods employing new evolution schedules in an analog
  computer with applications to determining isomorphic graphs and
  post-processing solutions, 2015.
\newblock Patent number: 20150363708.

\bibitem{rev_anneal_announce}
D-wave announces upgrades to d-wave 2000q quantum computer.
\newblock\href{http://www.marketwired.com/press-release/d-wave-announces-upgrades-to-d-wave-2000q-quantum-computer-2240453.htm}{http://www.marketwired.com/press-}
release/d-wave-announces-upgrades-to-d-
wave-2000q-quantum-computer-2240453.htm
\newblock Accessed: 2017-17-11.

\bibitem{Perdomo-Ortiz2011}
Alejandro Perdomo-Ortiz, Salvador~E. Venegas-Andraca, and Al{\'a}n
  Aspuru-Guzik.
\newblock A study of heuristic guesses for adiabatic quantum computation.
\newblock {\em Quantum Information Processing}, 10(1):33--52, 2011.

\bibitem{Ramezanpour2017}
A.~Ramezanpour.
\newblock Optimization by a quantum reinforcement algorithm.
\newblock {\em Phys. Rev. A}, 96:052307, Nov 2017.

\bibitem{macKaybookGA}
D.~J.~C. MacKay.
\newblock {\em Information Theory, Inference, and Learning Algorithms}.
\newblock Cambridge University Press, 2003.

\bibitem{Deng1999}
X.~Deng and P.~Fan.
\newblock New binary sequences with good aperiodic autocorrelations obtained by
  evolutionary algorithm.
\newblock {\em IEEE Communication Letters}, 3(10), 1999.

\bibitem{Fogel1994}
D.~B. Fogel.
\newblock An introduction to simulated evolutionary optimization.
\newblock {\em IEEE Transactions on Neural Networks}, 5(1), 1994.

\bibitem{Karimi2017}
Hamed Karimi and Gili Rosenberg.
\newblock Boosting quantum annealer performance via sample persistence.
\newblock {\em Quantum Information Processing}, 16(7):166, May 2017.

\bibitem{Karimi2017a}
Hamed Karimi, Gili Rosenberg, and Helmut~G. Katzgraber.
\newblock Effective optimization using sample persistence: A case study on
  quantum annealers and various monte carlo optimization methods.
\newblock {\em Phys. Rev. E}, 96:043312, Oct 2017.

\bibitem{Raymond2016}
J.~Raymond, S.~Yarkoni, and E.~Andriyash.
\newblock Global warming: Temperature estimation in annealers.
\newblock arXiv:quant-ph/1606.00919, 2016.

\bibitem{Sherrington1975}
D.~Sherrington and S.~Kirkpatrick.
\newblock Solvable model of a spin-glass.
\newblock {\em Phys. Rev. Lett.}, 35:1792--1796, Dec 1975.

\bibitem{Martonak2002}
R.~Marto\ifmmode \check{n}\else \v{n}\fi{}\'ak, G.~E. Santoro, and E.~Tosatti.
\newblock Quantum annealing by the path-integral monte carlo method: The
  two-dimensional random {I}sing model.
\newblock {\em Phys. Rev. B}, 66:094203, Sep 2002.

\bibitem{Nishimura2016}
Kohji Nishimura, Hidetoshi Nishimori, Andrew~J. Ochoa, and Helmut~G.
  Katzgraber.
\newblock Retrieving the ground state of spin glasses using thermal noise:
  Performance of quantum annealing at finite temperatures.
\newblock {\em Phys. Rev. E}, 94:032105, Sep 2016.

\bibitem{qbsolve}
A.~Douglass et. al.
\newblock qbsolve.
\newblock \url{https://github.com/dwavesystems/qbsolv}, accessed: Nov. 28,
  2017.

\bibitem{Bian2014}
Z.~Bian et. al.
\newblock Discrete optimization using quantum annealing on sparse {I}sing
  models.
\newblock {\em Frontiers in Physics}, 2(56), 2014.

\bibitem{Bian2016}
Z.~Bian, F.~Chudak, R.~B. Israel, B.~Lackey, W.~G. Macready, and A.~Roy.
\newblock Mapping constrained optimization problems to quantum annealing with
  application to fault diagnosis.
\newblock {\em Frontiers in ICT}, 3:14, 2016.

\bibitem{Swendsen1986}
R.~H. Swendsen and J.~S. Wang.
\newblock Replica monte carlo simulation of spin-glasses.
\newblock {\em Phys. Rev. Lett.}, 57:2607, 1968.

\bibitem{Earl2005}
D.J. Earl and M.~W. Deem.
\newblock Parallel tempering: Theory, applications, and new perspectives.
\newblock {\em Phys. Chem. Chem. Phys}, 7:3910--3916, 2005.

\bibitem{Hukushima2003}
K.~Hukushima and Y.~Iba.
\newblock {\em The Monte Carlo Method in the Physical Sciences: Celebrating the
  50th Anniversary of the {M}etropolis Algorithm}, volume 690.
\newblock AIP, 2003.

\bibitem{Matcha2010}
J.~Matcha.
\newblock Population annealing with weighted averages: A monte carlo method for
  rough free energy landscapes.
\newblock {\em Phys. Rev. E}, 82:026704, 2010.

\bibitem{Wang2015}
W.~Wang, J.~Machta, and H.~G. Katzgraber.
\newblock Population annealing: Theory and application in spin glasses.
\newblock {\em Phys. Rev. E}, 92:063307, 2015.

\bibitem{Barzegar2017}
A.~Barzegar, C.~Pattison, W.~Wang, and H.~G. Katzgraber.
\newblock Optimization of population annealing {M}onte {C}arlo for large-scale
  spin-glass simulations.
\newblock ar$\chi$iv:1710.09025, 2017.

\bibitem{LantingAQC2016}
T.~Lanting et. al.
\newblock Techniques for modifying annealing trajectories in quantum annealing
  processors.
\newblock https://aqccreg2016.eventfarm.com
  /events/index/7fff5387-0000-456c-a4da-3f0389a7aa72?page=7fff46cb-0000-4577-a218-60207dca65cd,
  2016.
\newblock presented at: AQC 2016.

\bibitem{Yedida2005}
J.~S. Yedidia, W.~T. Freeman, and Y.~Weiss.
\newblock Constructing free-energy approximations and generalized belief
  propagation algorithms.
\newblock {\em Information Theory, IEEE Transactionson}, 51(7):2282--2312,
  2005.

\bibitem{Kschischang2006}
F.~R. Kschischang, B.~J. Frey, and H.~A. Loeliger.
\newblock Factor graphs and the sum-product algorithm.
\newblock {\em IEEE Trans Inf Theor.}, 47:498--519, 2006.

\bibitem{Mceliece1998}
R.~J. Mceliece, D.~J.~C. Mackay, and J.~Cheng.
\newblock Turbo decoding as an instance of {P}earls belief propagation
  algorithm.
\newblock {\em IEEE J Select Areas Commun.}, 16:140--152, 1998.

\bibitem{Jarett2016}
M.~Jarret, S.~P. Jordan, and B.~Lackey.
\newblock Adiabatic optimization versus diffusion {M}onte {C}arlo methods.
\newblock {\em Phys. Rev. A}, 94:042318, Oct 2016.

\bibitem{Jarett2017}
B.~Lackey M.~Jarret.
\newblock Substochastic {M}onte {C}arlo algorithms.
\newblock arXiv:1704.09014, 2017.

\bibitem{Otsubo2012}
Y.~Otsubo et. al.
\newblock Effect of quantum fluctuation in error-correcting codes.
\newblock {\em Phys. Rev. E}, 86:051138, 2012.

\bibitem{Nishimori1980}
H.~Nishimori.
\newblock Exact results and critical properties of the {I}sing model with
  competing interactions.
\newblock {\em Journal of Physics C: Solid State Physics}, 13:4071, 1980.

\bibitem{Nishimori2001}
H.~Nishimori.
\newblock {\em Statistical Physics of Spin Glasses and Information Processing}.
\newblock Clarindon Press, 2001.

\bibitem{Frieden1972}
B.~R. Frieden.
\newblock Restoring with maximum likelihood and maximum entropy.
\newblock {\em Journal of the Optical Society of America}, 62:511, 1972.

\bibitem{Berger1996}
A.~L.~Berger et. al.
\newblock A maximum entropy approach to natural language processing.
\newblock {\em Computational Linguistics}, 22:39, 1996.

\bibitem{Phillips2006}
S.~J.~Phillips et. al.
\newblock Maximum entropy modeling of species geographic distributions.
\newblock {\em Ecological Modelling}, 190:231, 2006.

\bibitem{Raychaudhur2002}
S.~Raychaudhur et. al.
\newblock Associating genes with gene ontology codes using a maximum entropy
  analysis of biomedical literature.
\newblock {\em Genome Res.}, 12:203, 2002.

\bibitem{Gilmore1996}
C.~J. Gilmore.
\newblock Maximum entropy and {B}ayesian statistics in crystallography: A
  review of practical applications.
\newblock {\em Acta Crystallographica Section A}, 52:561, 1996.

\bibitem{Mistrulli2011}
P.~E. Mistrulli.
\newblock Assessing financial contagion in the interbank market: Maximum
  entropy versus observed interbank lending patterns.
\newblock {\em Journal of Banking and Finance}, 35:1114, 2011.

\bibitem{Rujan1993}
P.~Ruj\'an.
\newblock Finite temperature error-correcting codes.
\newblock {\em Phys. Rev. Lett.}, 70:2968--2971, May 1993.

\end{thebibliography}

\end{document}